Федушко С. С.

*Асистент кафедри соціальних комунікацій та інформаційної діяльності Національного університету «Львівська політехніка», Львів, Україна*

# РОЗРОБЛЕННЯ СИСТЕМИ ВЕРИФІКАЦІЇ СОЦІАЛЬНО-ДЕМОГРАФІЧНИХ ДАНИХ УЧАСНИКА ВІРТУАЛЬНОЇ СПІЛЬНОТИ

Вирішено важливе завдання розроблення системи верифікації соціально-демографічних даних учасника віртуальної спільноти на основі проведення комп'ютерно-лінгвістичного аналізу інформаційного наповнення великої вибірки україномовних віртуальних спільнот. Предметом дослідження є методи та засоби перевірки достовірності соціально-демографічних характеристик учасників віртуальних спільнот на основі комп'ютерно-лінгвістичного аналізу результатів їх комунікативної взаємодії. Метою роботи є перевірка правдивості персональних даних, які учасники надають у своїх облікових даних на основі результатів комп'ютерно-лінгвістичного аналізу інформаційних слідів учасників веб-спільнот. Для практичної реалізації поставлених завдань розроблено структуру програмного забезпечення для верифікації соціально-демографічного портрета веб-учасника. Запропоновано метод верифікації персональних даних учасника віртуальної спільноти на основі аналізу інформаційних слідів учасників віртуальних спільнот. Уперше розроблено метод перевірки достовірності персональних даних веб-учасників, що дало змогу спроектувати засіб верифікації соціально-демографічних характеристик учасників веб-спільноти. У результаті проведених досліджень розроблено систему верифікації соціально-демографічних даних учасників віртуальних спільнот, яка формує верифіковані соціально-демографічні портрети цих учасників віртуальних спільнот. Також представлено користувацький інтерфейс розробленої системи верифікації соціально-демографічних даних учасників віртуальних спільнот. Отримані результати системи дозволяють підвищити ефективність загального процесу управління веб-спільнотами. Апробації розроблених методів і засобів для вирішення завдань у веб-адмініструванні спільноти доводиться їх результативність та ефективність. Кількість фальшивих спрацювань системи верифікації не перевищує 18%.

**Ключові слова:** система верифікації, персональні дані, соціально-демографічні характеристики, аналіз, учасник, віртуальна спільнота.

**НОМЕНКЛАТУРА**

WWW – всесвітня мережа (англ. World Wide Web);
ВС – віртуальна спільнота;
ОЗ – обліковий запис;
СД – соціально-демографічний;
СДП – соціально-демографічний портрет;
СДХ – соціально-демографічні характеристика;
УВС – учасник віртуальної спільноти.

**ВСТУП**

Популярність віртуальних спільнот з кожним днем збільшується. Практично, таке інтерактивне спілкування у віртуальних спільнотах є невід'ємною частиною особистого та професійного життя сучасної людини.

Користувачі віртуальних спільнот надають про себе великі об'єми інформації, зокрема і конфіденційної.

Перевірка правдивості персональних даних, які учасники надають у своїх облікових даних є актуальним завданням у зв'язку з необхідністю виявлення протиправних дії веб-користувачів, які є образливими для опонентів з психологічного, фінансового, юридичного погляду. Зокрема, виявлення веб-користувачів, які займаються шантажем, надсилають листи з погрозами, поширюють брехливу і наклепницьку інформацію та займаються іншими видами електронного хуліганства, що полягає у проведенні комп'ютерно-лінгвістичного аналізу інформаційного наповнення доволі великої вибірки віртуальних спільнот Укрнету.

**1 ПОСТАНОВКА ЗАДАЧІ**

Вирішення актуальних завдань наукових досліджень у віртуальних спільнотах є необхідністю функціонування будь-якої віртуальної спільноти. Актуальні завдання сучасних наукових досліджень соціальних спільнот можна поділити на три типи: правові (вдосконалення системи правил комунікативної поведінки для УВС); організаційні (методи покращення функціонування системи віртуальних спільнот сприяє підвищенню ефективності та покращення функціонування веб-спільнот); економічні (розроблення нових механізмів та підходів до застосування інтернет-реклами, таргетингу).

Підвищення ефективності функціонування ВС – результат виконання таких завдань:

– відсіювання небажаного інформаційного наповнення;
– підвищення якості інформаційного наповнення;
– фільтрація учасників за повнотою заповнення ОЗ персональними даними;
– зменшення витрат на модерування спільнотою;
– зменшення конфліктних ситуацій у спільноті;
– комп'ютерно-лінгвістичний аналіз достовірності персональних даних УВС;
– автоматизування комп'ютерно-лінгвістичного опрацювання контенту.

Метою роботи є розроблення методів і засобів перевірки достовірності СДХ учасників віртуальних спільнот за результатами комп'ютерно-лінгвістичного аналізу інформаційних слідів учасників ВС.

Мета роботи визначає необхідність виконання таких актуальних задач:

– дослідити специфіку та виділити основні ознаки онлайн-комунікації УВС шляхом комп'ютерно-лінгвістичного аналізу їхнього інформаційного наповнення;
– спроектувати структуру програмного комплексу комп'ютерно-лінгвістичної перевірки достовірності СДХ учасника віртуальної спільноти;
– створити засіб верифікації персональних даних учасників веб-спільноти;
– апробувати програмний комплекс комп'ютерно-лінгвістичної перевірки достовірності соціально-демографічних характеристик УВС.

Отже, реалізація наведених вище завдань вагомо впливає на функціонування ВС, що дає можливість спростити та пришвидшити виконання обов'язків адміністратора ВС.

## 2 ОГЛЯД ЛІТЕРАТУРИ

У дослідженнях віртуальних спільнот у WWW науковці виокремлюють чимало напрямів, оскільки у зв'язку з появою нових тенденцій розвитку віртуальних спільнот сфера досліджень збільшується буквально щодня.

Проаналізувавши сучасні праці науковців [1–5], виявлено, що в останні роки з'явилося багато публікацій про особливості функціонування віртуальних спільнот.

Дослідження віртуальних спільнот почались з вивчення комп'ютерно-опосередкованої комунікації – комунікації, яка ґрунтується на використанні комп'ютера, що спрощує та розширює сферу комунікації [2] та дослідженням соціальної зміни пов'язана відчуження і втратою спільноти у кіберпросторі, створюючи спільноти за інтересами [3], спричиненої впровадженням комп'ютерних технологій. Поняття та функціонування віртуальних спільнот є дискусійним питання для багатьох науковців [3].

Тематичні конференції (веб-форуми та дискусійні групи) в мережі Інтернет [1], електронні дошки оголошень та інші види комп'ютерно-опосередкованої комунікації виникли з метою відтворити відчуття спільноти, що з'являється в учасників ВС, які беруть участь у комунікації з метою відновлення соціальних зв'язків у віртуальній спільноті.

Компонента спільноти як зв'язку певних груп, має найбільше значення в контексті комп'ютерно-опосередкованої комунікації і твердить, що ці три ознаки є в комп'ютерно-опосередкованій комунікації.

Віртуальна спільнота [4] – це група людей зі спільними інтересами, регулярною та тривалою комунікацією за допомогою Інтернету через спільне місцезнаходження або механізм. З іншого боку, віртуальну спільноту визначають як мережу взаємозв'язків осіб у кіберпросторі [5]. Часто веб-спільноти суттєво впливають на їхніх учасників, пропонуючи можливості навчального середовища, досягнення інтерактивності та способи обміну архівів контактів у системах знань.

Аналіз діяльності віртуальних спільнот є об'єктом наукових досліджень, серед яких чітко виділяються три основні напрями: аналіз дій учасників (web usage mining); аналіз веб-структур (web structure mining); аналіз контенту (web content mining).

Однією із важливих проблем аналізу контенту ВС сьогодні є аналіз персональних даних ВС. Попри суттєве значення для подальшого розвитку цієї сфери досліджень досі не розроблено дієвих методів аналізу саме персональної інформації в обліковому записі УВС.

Зважаючи на вищенаведений аналіз наукових праць, серед відомих літературних джерел відчутним є брак ґрунтовних досліджень з верифікації персональної інформації користувачів мережі Інтернет та дослідження достовірності СДХ учасників соціальних комунікацій, зокрема ВС з метою покращення їх функціонування. Це, своєю чергою, породжує актуальну проблему розроблення нових методів та засобів аналізу достовірності СДХ учасників віртуальних спільнот, які би мали належне наукове обґрунтування, формалізацію, прогнозовану результативність та універсальність.

## 3 МАТЕРІАЛИ ТА МЕТОДИ

Ефективне функціонування спільноти залежить від ряду чинників, як об'єктивних, так і суб'єктивних. Проте, наявні на сьогодні методи та засоби, що використовуються у глобальних сервісах, в повній мірі не виконують усіх завдань, які б задовольнили потреби власників та модераторів в управління віртуальними проектами, зокрема віртуальними спільнотами. Розроблення методів та засобів перевірки достовірності персональних даних у обліковому записі користувача є одним з найбільш значимих факторів, які впливають на покращення функціонування віртуальної спільноти.

Дослідження, виконані під час роботи, ґрунтуються на методах структурного аналізу для дослідження інформаційного наповнення веб-учасників та методи аналізу веб-контенту, спрямовані на дослідження вербальних характеристик текстів. Для вирішення завдань моделювання структури системи валідації СД портрета веб-учасника використано теоретико-множинні підходи та апарат теорії реляційних баз даних. Засоби формування соціально-демографічних портретів веб-учасників розроблено на основі сучасних веб-технологій.

До програмного засобу для верифікації СДП веб-учасника ставляться високі вимоги, адже цей засіб в реальному часі повинен опрацьовувати великий масив динамічної інформації. Також треба звертати увагу на постійний ріст активності учасників популярної ВС, які невпинно генерують не завжди якісне та адекватне, до цілей позиціонування спільноти у вебі, інформаційне наповнення. Звичайно, такий сценарій розвитку ВС є перешкодою для ефективного функціонування спільноти в конкурентному середовищі.

Серед задач, які вимагають окремого підходу до валідації персональних даних УВС, методом комп'ютерно-лінгвістичного аналізу контенту, виокремлено наступні:

– брак якісного та достовірного інформаційного наповнення;

– мінімалізація анонімності користувачів [6–7];

– сприйняття спільноти, як платформи для довільної реалізації своїх бізнес ідей [8–9];

– невелика успішність ВС, що призводить до низької рентабельності веб-проекту;

– необхідність великих капіталовкладень власниками веб-спільноти;

– вагомі та необґрунтовані затрати часу у модеруванні спільнотою;

– уникнення загроз інформаційної безпеки учасників ВС, що можуть призвести до кримінальної відповідальності власника спільноти [10];

– низька конкурентоспроможність в порівнянні з іншими спільнотами;

– проблеми в управлінні віртуальною спільнотою (високий рівень конфліктів, низький авторитет модераторів та ін.);

– покращення методів інтернет-таргетингу [11–12].

Усі вищеперераховані задачі враховані при розробленні структури програмного забезпечення для верифікації *соціально-демографічного портрета веб-учасника* [13–16]. Розроблення саме програмного засобу верифікації – «Верифікатор СДХ веб-учасника» (див. рис. 1) вирішить багато актуальних задач в модеруванні та управлінні спільнотами у WWW.

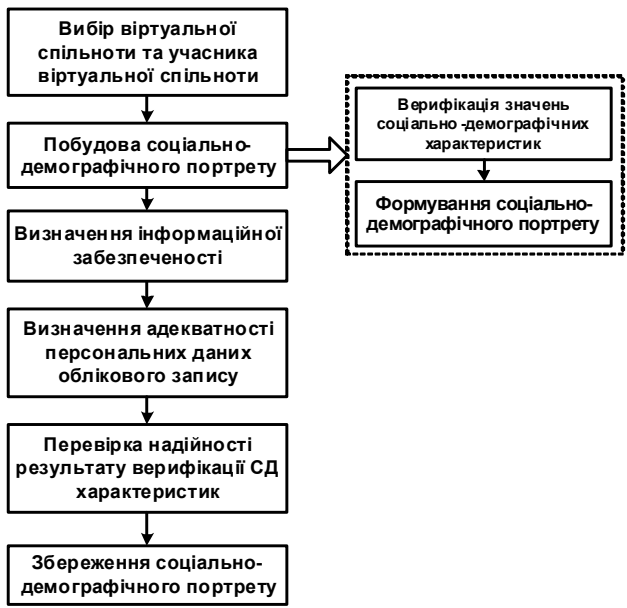

Рисунок 1 – Схема функціонування системи
«Верифікатор СДХ веб-учасника»

Для верифікації СДХ та побудови соціально-демографічного портрета УВС автором розроблено чітку структуру програмного забезпечення «Верифікатор соціально-демографічних характеристик веб-учасника». Ключовим елементом цієї структури є схема функціонування цієї системи. Тепер детально опишемо кожен етап схеми функціонування системи «Верифікатор соціально-демографічних характеристик веб-учасника».

З переліку віртуальних спільнот, які належать власнику чи доступні модератору, здійснюється вибір віртуальної спільноти і з бази учасників обраної віртуальної спільноти обираємо конкретного учасника або множини учасників.

Соціально-демографічний портрет будуємо лише з верифікованих СДХ учасника ВС, тобто з персональних даних учасника, достовірність яких перевірено методом комп'ютерно-лінгвістичного аналізу. Тож, побудова СД портрета ґрунтується на верифікації СДХ учасника ВС. Побудова СДП відбуває згідно алгоритму формування СДП учасника віртуальної спільноти. Блок-схема алгоритму формування СДП зображено на рис. 2.

Мета алгоритму – перевірити достовірність максимальної кількості персональної інформації, яку УВС вказав у своєму обліковому записи методом комп'ютерно-лінгвістичної верифікації СДХ учасника ВС.

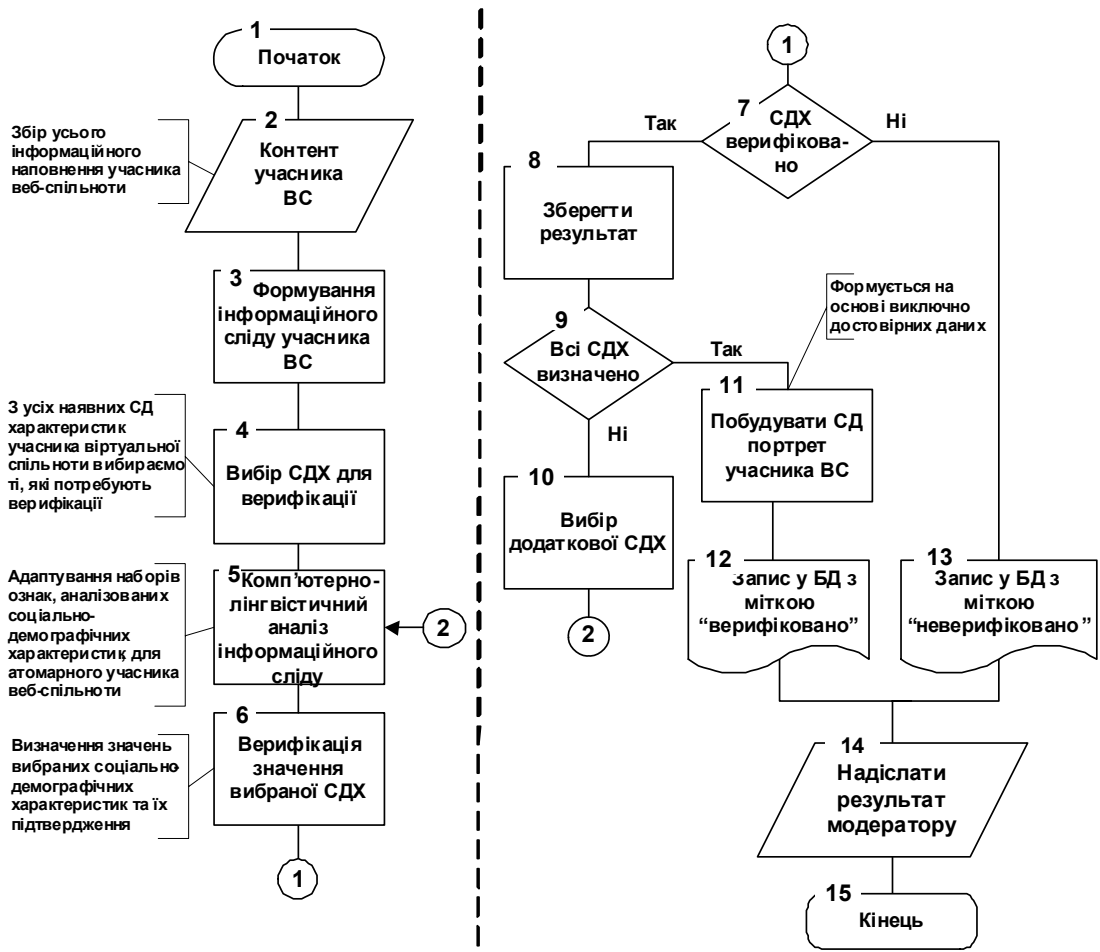

Рисунок 2 – Блок-схема алгоритму формування СДП учасника віртуальної спільноти

Основою алгоритму є інформаційний слід УВС – інформаційного наповнення віртуальної спільноти конкретного УВС.

Відповідно до цього алгоритму, є можливість здати параметри вибору характеристик, тобто обрати саме потрібні СДХ для верифікації. Перевірку достовірності всіх СД характеристик налаштовано по замовчуванню.

Соціально-демографічний портрет УВС будується виключно з верифікованих СДХ. Тобто відповідність вказаним даним у обліковому записі реальним характеристикам УВС. На основі верифікації СДХ та побудованого СДП система «Верифікатор СДХ веб-учасника» класифікує учасників за фактом здійснення чи нездійснення перевірки достовірності.

*Формування соціально-демографічного портрета.* На цьому етапі відбувається саме вже формування СДП на основі результатів верифікації СДХ. Система формує результати у табличному вигляді.

### 4 ЕКСПЕРИМЕНТИ

Для вирішення вищенаведених завдань потрібно спроектувати програмний засіб для верифікації персональних даних УВС. Зважаючи на специфіку середовищ соціальних комунікацій, варто звернути увагу на деякі спеціальні вимоги до програмного засобу «Верифікатор СДХ веб-учасника»: задоволення потреб користувача при розв'язанні конкретної функціональної задачі; зрозумілість і логічність виводу результатів, що зменшить час прийняття рішення та покращить їх якість; можливість нарощення функціональності, тобто адаптування та збільшення функціонального блоку програмного засобу відповідно до потреб конкретного учасника [16]; простота аналізу, збереження та використання результатів, впровадження їх у діяльність віртуальної спільноти. Відповідно до цих вимог створено програмне забезпечення для верифікації СДП веб-учасника (рис. 3).

Результати програми «Верифікатор соціально-демографічних характеристик веб-учасника» з результатами верифікації даних учасника веб-форуму «Львів. Форум Рідного Міста» – Andreas представлено на рис. 4.

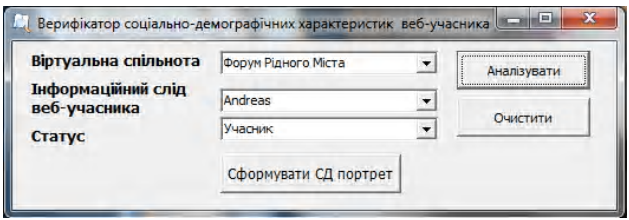

Рисунок 3 – Запит на верифікацію облікового запису веб-форуму «Львів. Форум Рідного Міста» користувача Andreas

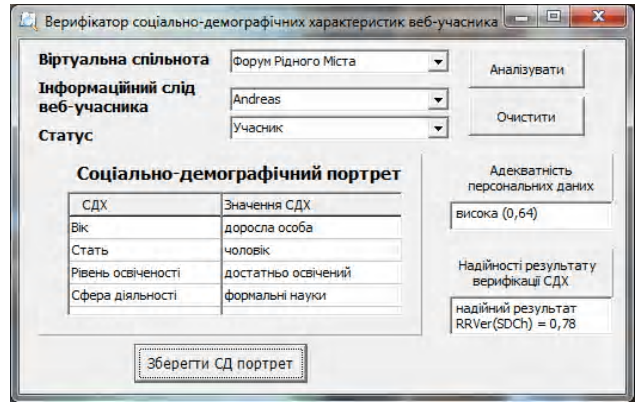

Рисунок 4 – Результати верифікації СДП учасника «Львів. Форум Рідного Міста» – Andreas

Модель словника СД маркерів є складовою цієї системи «Верифікатор соціально-демографічних характеристик веб-учасника».

Для ефективного використання програмного засобу «Верифікатор соціально-демографічних характеристик веб-учасника» результати його роботи автоматично імплементовані у функціонування ВС.

Апробація такого методу здійснено у таких віртуальних спільнотах [17–18]: «Львів. Форум Рідного Міста», «Малеча», «Дівочі посиденьки», «Rock.Lviv.Ua» та «Теревені».

### 5 РЕЗУЛЬТАТИ

За алгоритмом обчислення надійності результату процесу верифікації СДХ визначено кількість фальшивих спрацьовувань системи верифікації СДХ учасників п'ятьох досліджуваних віртуальних спільнот. Результати подано у табличному варіанті (див. табл. 1).

### 6 ОБГОВОРЕННЯ

На сьогоднішній день для задач верифікації інформації УВС існують лише дорогі комплекси рішення, які не задовольняють усіх потреб власників в задачі управління ВС. Розроблення якісного та відносно недорогого механізму валідації СДХ зумовлене потребою в ідентифікації користувачів на етапі бурхливого розвитку віртуальної спільноти.

Відносна кількість фальшивих спрацьовувань системи верифікації персональних даних УВС на тестовій множині учасників п'яти популярних веб-спільнот прийнятна для завдань управління ВС і фактично є ймовірністю помилки системи (табл. 1).

Ефективність системи верифікації персональних даних [19] показує, що у 2–4 рази (залежно від типу, мети та специфіки веб-спільноти) зменшується навантаження з перевірки персональних даних на модераторів, відповідно і суттєвим є зниження часових та фінансових затрат на адміністрування віртуальних спільнот.

Таблиця 1 – Результативність системи верифікації СД характеристик

| Віртуальні спільноти | Малеча | Дівочі посиденьки | Rock.Lviv.Ua | Теревені | Львів. Форум Рідного Міста |
|---|---|---|---|---|---|
| Кількість перевірених учасників | 1631 | 504 | 216 | 386 | 345 |
| Відносна кількість фальшивих спрацьовувань, % | 13 | 8 | 16 | 18 | 7 |
| Ефективність, % | 70 | 62 | 55 | 58 | 62 |

**ВИСНОВКИ**

У роботі вирішено актуальну задачу розроблення системи верифікації інформації учасників ВС, на основі комп'ютерно-лінгвістичних методів верифікації масиву користувацького контенту ВС.

Наукова новизна одержаних результатів полягає у науковому обгрунтуванні та розробленні методів та засобів верифікації СДХ учасників веб-спільнот.

Уперше розроблено метод перевірки достовірності персональних даних ВС, який полягає у порівнянні заявлених СДХ у облікових записах ВС із їхнього СДП, які сформовані на основі результатів комп'ютерно-лінгвістичного аналізу інформаційних слідів ВС, що дало змогу спроектувати засіб верифікації СДХ учасника веб-спільноти. Удосконалено метод адміністрування ВС, що дало змогу істотно зменшити кількість недостовірного контенту ВС та виявляти неправдиві персональні дані веб-учасників.

Розроблено метод комп'ютерно-лінгвістичного аналізу інформаційного сліду ВС шляхом відбору та аналізу особистісних лексико-семантичних та лексико-синтаксичних, граматичних особливостей мовлення кожного окремого ВС та побудови його СДП, що дає змогу верифікувати персональні дані, які задекларував цей УВС.

Практичне значення одержаних результатів роботи зумовлено тим, що вони дають змогу перевірити достовірність персональних даних УВС на основі аналізу їхніх інформаційних слідів.

Розроблено алгоритми верифікації персональних даних облікових записів УВС, шляхом побудови їхніх інформаційних слідів на основі результатів аналізу контенту ВС, що дало змогу підвищити їхню керованість та ефективність адміністрування.

Розроблено структуру програмного комплексу засобів перевірки достовірності даних шляхом формування СДП для автоматизації процесу верифікації масиву користувацького інформаційного наповнення.

Кількість фальшивих спрацювань системи верифікації СДХ учасника веб-спільноти не перевищує 18%.

Таким чином, результати апробації комплексу комп'ютерно-лінгвістичної перевірки достовірності СДХ учасника веб-спільноти показали, що запропонований метод перевірки достовірності соціально-демографічних характеристик УВС на основі методів верифікації масиву користувацького інформаційного наповнення WWW ефективніший від відомих методів, і дало змогу реалізувати систему підтримки прийняття рішень для фахівця з управління веб-спільнотами у сфері опрацювання персональних даних.

Перспективи подальших досліджень полягають у застосуванні запропонованого підходу для вирішення практичних задач управління віртуальними спільнотами.

Федушко С. С.

Ассистент кафедры социальных коммуникаций и информационной деятельности Национального университета «Львовская политехника», Львов, Украина

**РАЗРАБОТКА СИСТЕМЫ ВЕРИФИКАЦИИ СОЦИАЛЬНО-ДЕМОГРАФИЧЕСКИХ ДАННЫХ УЧАСТНИКОВ ВИРТУАЛЬНЫХ СООБЩЕСТВ**

Решено важная задача разработки системы верификации социально-демографических данных участника виртуального сообщества на основе проведения компьютерно-лингвистического анализа информационного наполнения большой выборки украиноязычных виртуальных сообществ. Предметом исследования являются методы и средства проверки подлинности социально-демографических характеристик участников виртуальных сообществ на основе компьютерно-лингвистического анализа результатов их коммуникативного взаимодействия. Целью работы является проверка достоверности персональных данных, которые участники предоставляют в своих учетных данных на основе результатов компьютерно-лингвистического анализа информационных следов участников веб-сообществ. Для практической реализации поставленных задач разработана структура программного обеспечения для верификации социально-демографического портрета веб-участника. Предложен метод верификации персональных данных участника виртуального сообщества на основе анализа информационных следов участников виртуальных сообществ. Впервые разработан метод проверки достоверности персональных данных о участников, что позволило спроектировать средство верификации социально-демографических характеристик участника веб-сообщества. В результате проведенных исследований разработана система верификации социально-демографических данных участников виртуальных сообществ, которая формирует верифицированы социально-демографические портреты этих участников виртуальных сообществ. Также представлены интерфейс разработанной системы верификации социально-демографических данных участников виртуальных сообществ. Полученные результаты системы позволяют повысить эффективность общего процесса управления веб-сообществами. Апробации разработанных методов и средств для решения задач в веб-администрирования сообщества приходится их результативность и эффективность. Количество фальшивых срабатываний системы верификации не превышает 18%.

**Ключевые слова:** система верификации, персональные данные, социально-демографические характеристики, анализ, участник, виртуальная сообщество.

Fedushko S. S.

Assistant of Social Communications and Information Activities Department of Lviv Polytechnic National University, Lviv, Ukraine

**DEVELOPMENT OF VERIFICATION SYSTEM OF SOCIO-DEMOGRAPHIC DATA OF VIRTUAL COMMUNITY MEMBER**

The important task of developing verification system of data of virtual community member on the basis of computer -linguistic analysis of the content of a large sample of Ukrainian virtual communities is solved. The subject of research is methods and tools for verification of web-members socio-demographic characteristics based on computer-linguistic analysis of their communicative interaction results. The aim of paper is to verifying web-user personal data on the basis of computer-linguistic analysis of web-members information tracks. The structure of verification software for web-user profile is designed for a practical implementation of assigned tasks. The method of personal data verification of web-members by analyzing information track of virtual community member is conducted. For the first time the method for checking the authenticity of web members personal data, which helped to design of verification tool for socio-demographic characteristics of web-member is developed. The verification system of data of web-members, which forms the verified socio-demographic profiles of web-members, is developed as a result of conducted experiments. Also the user interface of the developed verification system web-members data is presented. Effectiveness and efficiency of use of the developed methods and means for solving tasks in web-communities administration is proved by their approbation. The number of false results of verification system is 18%.

**Keywords:** system of verification, personal data, socio-demographic characteristics, analysis, member, virtual community.